\documentclass[runningheads]{llncs}
\usepackage{graphicx}
\usepackage{booktabs} 
\usepackage{multirow}
\usepackage{cite}
\usepackage[hidelinks]{hyperref}
\usepackage{xcolor,soul}
\usepackage{url}
\usepackage{bbding}
\usepackage{array}

\DeclareMathAlphabet{\mathpzc}{OT1}{pzc}{m}{it}

\usepackage{listings}
    \lstdefinestyle{mystyle}{
        frame=single,
        float = th,
        floatplacement=tbp,
        commentstyle=\color{codegreen},
        keywordstyle=\color{blue}\bfseries,
        numberstyle=\tiny\color{codegray},
        stringstyle=\color{codepurple},
        basicstyle=\linespread{0.5}\fontsize{8}{11.8}\ttfamily\bfseries,
        breakatwhitespace=false,
        breaklines=false,
        captionpos=b,
        keepspaces=true,
        numbers=none,
        showspaces=false,
        showstringspaces=false,
        showtabs=false,          
        tabsize=2,
        language=C,
        moredelim=**[is][\underbar]{_}{_},
        escapechar=\@,
        emptylines=1,
        numbers=left,
        numberstyle=\tiny\sffamily,
        numbersep=4pt,
        aboveskip=4pt,
        xleftmargin=2pt,
        frame=none,
        escapeinside={(*@}{@*)}
    }
\DeclareRobustCommand{\hlcyan}[1]{{\sethlcolor{yellow}\hl{#1}}}
\DeclareRobustCommand{\hlcsec}[1]{{\sethlcolor{red}\hl{#1}}}
\usepackage{xcolor}
\usepackage[shortlabels]{enumitem}

\let\oldmaketitle\maketitle
\renewcommand{\maketitle}{\oldmaketitle\setcounter{footnote}{0}}

\begin{document}
\title{Enhancing Pre-Trained Language Models for Vulnerability Detection via Semantic-Preserving Data Augmentation}
\titlerunning{Enhancing Pre-Trained Language Models for Vulnerability Detection}
%
\author{Weiliang Qi~\inst{1}, Jiahao Cao~\inst{2}, Darsh Poddar~\inst{3}, Sophia Li~\inst{4}, Xinda Wang~\inst{1}}
\authorrunning{W. Qi et al.}
%
\institute{University of Texas at Dallas \and Tsinghua University \and
Lebanon Trail High School \and
Lovejoy High School \\
\email{\{weiliang.qi, xinda.wang\}@utdallas.edu}, 
\\\email{caojh2021@tsinghua.edu.cn},
\\\email{darsh.pod@gmail.com}, \email{sophia.li2025@gmail.com}}
\maketitle              

\begin{abstract}
With the rapid development and widespread use of advanced network systems, software vulnerabilities pose a significant threat to secure communications and networking.
Learning-based vulnerability detection systems, particularly those leveraging pre-trained language models, have demonstrated significant potential in promptly identifying vulnerabilities in communication networks and reducing the risk of exploitation.
However, the shortage of accurately labeled vulnerability datasets hinders further progress in this field. 
Failing to represent real-world vulnerability data variety and preserve vulnerability semantics, existing augmentation approaches provide limited or even counterproductive contributions to model training.

In this paper, we propose a data augmentation technique aimed at enhancing the performance of pre-trained language models for vulnerability detection.
Given the vulnerability dataset, our method performs natural semantic-preserving program transformation to generate a large volume of new samples with enriched data diversity and variety.
By incorporating our augmented dataset in fine-tuning a series of representative code pre-trained models (i.e., CodeBERT, GraphCodeBERT, UnixCoder, and PDBERT), up to 10.1\% increase in accuracy and 23.6\% increase in F1 can be achieved in the vulnerability detection task. 
Comparison results also show that our proposed method can substantially outperform other prominent vulnerability augmentation approaches.

\keywords{Vulnerability Detection  \and Language Model \and Data Augmentation.}
\end{abstract}
\section{Introduction}

The rapid growth of communications and networking has led to the wide deployment of numerous related software and systems. Meanwhile, we observe an increasing number of high-severity software vulnerabilities in these systems. The widely adopted secure communication toolkit, OpenSSL, was compromised by Heartbleed, where adversaries exploit the vulnerability remotely and steal private information from the server's memory~\cite{durumeric2014matter}, affecting millions of well-known websites including Google and Yahoo. 
The Equifax data breach is due to a vulnerability in the famous web application framework Apache Struts~\cite{struts}, resulting in the exposure of 148M SSNs. Nowadays, software vulnerabilities present a significant threat to secure communications and networking.



Learning-based vulnerability detection techniques can automatically identify vulnerabilities in communication network systems, enabling timely actions to defend against potential attacks. Among them, fine-tuning pre-trained language models for vulnerability detection tasks has presented the most promising effectiveness~\cite{steenhoek2023empirical}.
However, while fine-tuning requires much less data compared to the initial pre-training phase, there are still not enough vulnerability samples available to train a robust model~\cite{fan2020ac, zhou2019devign}.


To address the vulnerable data shortage, a series of works focus on the augmentation of the vulnerability datasets. 
SARD~\cite{SARD} generates synthetic vulnerability data by adding additional code elements that preserve the original program functionalities. Although this process adds more variety, it overlooks if the code still appears realistic, leading to poor test performance on real-world data
~\cite{chakraborty2021deep}. Another line of works~\cite{nong2024vgx, nong2023vulgen, zhang2022fixreverter} tries to augment the datasets through injecting the vulnerability into the non-vulnerable codebase. Generally, these works first generate some vulnerability patterns, and then search for positions in code snippets to insert these patterns. However, such approaches can only inject specific types of simple vulnerabilities. 
Moreover, the injection process may fail to consider the context of the vulnerability, making the injected vulnerability ineffective. Therefore, we need a data augmentation method that not only enriches the realistic diversity and variety of vulnerabilities but also retains their essential vulnerability characteristics.


Aiming to enhance the performance of pre-trained language models in vulnerability detection, this paper proposes a natural semantic-preserving data augmentation method to obtain a large number of vulnerability samples at a low cost. 
Specifically, our method performs transformations on existing code data in the vulnerability dataset to generate more samples. To ensure the generated samples are useful for model training, we designed our transformation to meet two specific requirements:
1) The transformation rules should align with human programming styles, making our generated data close to real-world vulnerability samples and helping models achieve better performance in practice. To this end, we craft seven natural transformation rules based on our observation of real-world samples. 2) The transformation needs to be semantic-preserving. 
This ensures that our transformations do not disrupt the semantics of the original vulnerabilities, thereby maintaining consistent vulnerability characteristics between the new data and the original data. 
Besides, semantic-preserving transformations enable us to apply transformation rules multiple times to the same code, targeting different applicable locations, and thereby generating a large number of diverse samples in a short period.
By augmenting the Big-Vul, a real-world vulnerability dataset collected from NVD~\cite{fan2020ac} that includes 188,636 samples, we generate 3,233,635 new samples, increasing the dataset size by 17 times.


We conduct an extensive evaluation on the effectiveness enhancement brought by our vulnerability augmentation. First, we apply our augmented dataset on fine-tuning four state-of-the-art (SOTA) code pre-trained language models (i.e., CodeBERT~\cite{feng2020codebert}, GraphCodeBERT~\cite{guo2020graphcodebert},
UnixCoder~\cite{guo2022unixcoder}, and PDBERT~\cite{liu2024pre}). Experimental results show that our augmented dataset is able to help achieve higher performance on all four models. Specifically,
on CodeBERT and GraphCodeBERT, our augmentation facilitates 8.7\% and 10.1\% accuracy increase as well as 15.5\% and 23.6\% F1 increase. 
Second, by comparing with existing prominent vulnerability augmentation approaches (i.e., SARD~\cite{SARD} and VGX~\cite{nong2024vgx}), our proposed method significantly outperform them with 14.0\% higher accuracy and 51.1\% higher F1 score.
Additionally, we also study the performance improvement by our augmentation on the top 25 frequent vulnerabilities in the real world. Results show that our method helps achieve a more than 30\% average increase on true positive rate.


In summary, we make the following contributions:

\begin{itemize}
\item We propose a new method to tackle the challenge of insufficient sample numbers in vulnerability datasets by employing semantic-preserving transformations for dataset augmentation. The source code of this work is available at \url{https://github.com/XSecLab-UTD/VDDA}.

\item We introduce seven natural semantic-preserving transformation rules and develop a toolset for the automatic augmentation of vulnerability datasets. We demonstrate their effectiveness on a widely adopted real-world vulnerability dataset.

\item To the best of our knowledge, we are the first to evaluate the program transformation-based data augmentation techniques on pre-trained language models. Experimental results show that our method greatly outperforms other vulnerability augmentation approaches and substantially improves the performance of pre-trained models on vulnerability detection tasks.


\end{itemize}


The rest of this paper is organized as follows. Section 2 provides preliminaries including our work's connection to secure communication and networking. Section 3 presents the details of our proposed semantic-preserving data augmentation method. In Section 4, we conduct extensive evaluations on the effectiveness of our work. In Section 5, we discuss the limitations of our approach and future work. Section 6 summarizes the related work and Section 7 concludes the paper.
\section{Preliminaries}



\subsection{Software Vulnerability in Communications and Networking}

Software vulnerabilities are a type of defect that attackers can exploit to threaten the security of the target system, such as gaining unauthorized access, terminating services, or accessing private information. Over the past decade, according to records from the National Vulnerability Database (NVD), the number of vulnerabilities that could be exploited from across a network\footnote{We compute the number of network-related vulnerabilities by checking the Attack Vector (AV) in each CVE entry. If the value of AV is labeled as ``Network (N)", such a vulnerability is termed “remotely exploitable” and considered a network attack being exploitable at the protocol level one or more network hops away, and therefore warrants a greater severity compared with the vulnerabilities requiring local/physical access to a device~\cite{cvss}.} (and consequently the resulting higher severity) has increased from 1,180 in 2013 to 19,231 in 2023, which poses a huge threat to secure communications and networking. 


On the other hand, as computer networks have become more advanced, more software and systems have been developed and employed, resulting in an increasing number of vulnerabilities. 
We calculate and present the top 10 software with the most Common Vulnerabilities and Exposures (CVEs) in \autoref{tab:bigvulprojects}, using datasets collected by Big-Vul~\cite{fan2020ac} from NVD. Among these software, 8 out of 10 are related to computer network applications, frameworks, or infrastructure. Due to the widespread use of computer network-related software and the huge number of vulnerabilities they possess, it is crucial to address these vulnerabilities to ensure security and privacy in communication networks.


\begin{table*}[htbp]
\vspace{-2mm}
\begin{center}
\setlength{\tabcolsep}{2mm}
\caption{Top 10 projects with the highest number of CVEs}
\vspace{-2mm}
\label{tab:bigvulprojects}
\footnotesize	
\begin{tabular}{c| c |c}
\toprule

\textbf{Project} & \textbf{Description \& Network-related (Y/N)} &  \textbf{\%}  \\
\hline\hline
     Chrome       &  Web browser (Y)          & 38.45$\%$  \\
\hline
     Linux        &  Operating system including networking subsystem (Y)            & 17.01$\%$  \\
\hline
     Android      &  Operating system including networking subsystem (Y)            & 11.96$\%$  \\
\hline
     ImageMagick  &  Image editor        & 2.91$\%$  \\
\hline
     OpenSSL      &   Secure communication Toolkit (Y)         & 1.54$\%$  \\
\hline
     tcpdump      &   Network packet analyzer (Y)         & 1.36$\%$   \\
\hline
     qemu         &     Machine emulator and virtualizer    & 1.20$\%$  \\
\hline
     PHP          &   Scripting language suited to web development (Y)        & 1.18$\%$  \\
\hline
     savannah     &  Online Software forge (Y)       & 1.08$\%$    \\
\hline
     krb5         &   Network authentication protocol (Y)         & 1.07$\%$  \\
\hline
\bottomrule
\end{tabular}
\end{center}
\end{table*}


\subsection{Pre-Trained Models for Vulnerability Detection}

To guarantee secure communications and networking, vulnerability detection plays an important role in discovering and prioritizing vulnerabilities so that they can be addressed through patches, configuration changes, or other security measures to reduce the risk of exploitation.

Traditional vulnerability detection approaches heavily rely on manual code reviews. To relieve human efforts, rule-based approaches initially involve human security experts summarizing a set of vulnerability patterns, followed by the development of tools that automate vulnerability detection by matching these patterns~\cite{liu2012software, pistoia2005interprocedural, pistoia2007role, cuoq2009experience}. 
Inspired by recent research advances in deep learning, a series of DL-based vulnerability detection techniques have been proposed and achieved promising results, which further reduce the human workload. Early works such as VulDeePecker~\cite{li2018vuldeepecker} and SySeVR~\cite{li2021sysevr} adopt Recurrent Neural Networks (RNNs) to automatically learn features from the source code, eliminating the need for manually defining patterns or features to predict the presence of vulnerabilities. 
Since program code contains rich structural information (e.g., control flow and data dependency) that can be effectively represented by graphs, subsequent works have applied Graph Neural Networks (GNNs). For instance, Devign~\cite{zhou2019devign} and ReVeal~\cite{chakraborty2021deep} first construct Code Property Graphs (CPGs) from the input code and then use GNN models for vulnerability detection.


With the recent development of the transformer with encoder-decoder architecture~\cite{vaswani2017attention}, most advanced works utilize the capabilities of pre-trained language models and fine-tune them for vulnerability detection tasks. 
These pre-trained models are trained on a large amount of code-related data, enabling them to better capture the code syntax and semantic features. Both the encoder and the decoder are built with multiple stacked layers that contain multi-head attention calculation and allocation mechanisms. The encoder focuses on generating the context vectors used by the model, while the decoder considers the context vectors of the model's output. Currently, popular pre-trained models can be categorized into three types based on their encoder-decoder architecture: one type only has the encoder structure~\cite{feng2020codebert}; another type only has the decoder~\cite{brown2020language, lu2021codexglue}; and the third type uses both encoder and decoder~\cite{wang2021codet5, guo2022unixcoder}. During the training process, these models use different code representations to enhance training effectiveness and help the model better capture code features. For example, CodeBERT~\cite{feng2020codebert} uses bimodal data pairing natural language and code to enhance pre-training, while UnixCoder~\cite{guo2022unixcoder} introduces abstract syntax tree (AST) information during pre-training, and GraphCodeBERT~\cite{guo2020graphcodebert} incorporates data flow information. 

Since pre-trained language models achieve SOTA performance on vulnerability detection~\cite{steenhoek2023empirical}, our work focuses on them and explores ways to further improve their effectiveness. 


\subsection{Data Augmentation for Vulnerability Detection}

Although fine-tuning a pre-trained language model for vulnerability detection needs less data compared to the initial pre-training phase, existing publicly accessible vulnerability datasets still fall short of meeting this requirement.
Unlike computer vision and natural language processing tasks, collecting high-quality large-scale vulnerability data is challenging in practice. Current popular vulnerability datasets~\cite{fan2020ac, chakraborty2021deep, zhou2019devign, wang2021patchdb} contain only around 10,000 vulnerability samples in C/C++ collected from GitHub repositories. In this case, data augmentation is a promising direction to address the limitation of real-world open-source vulnerabilities. 
A series of works attempt to transform real-world code snippets to generate new samples of vulnerability data. Some of them~\cite{zhang2022fixreverter, nong2023vulgen, nong2024vgx} try to inject vulnerability patterns on the benign code of the real world. Others~\cite{liu2024enhancing, lee2022fuzzle} add varieties on vulnerable code to generate new samples. However, these vulnerability augmentation techniques have two main limitations:

\textbf{Limitation 1}: Although synthetic datasets like SARD~\cite{SARD} and SATE IV~\cite{okun2013report} use semantic-preserving transformation to generate new data, they overlook the naturalness of the code. This results in a gap between the generated samples and real-world samples, which can limit the generalization capability of the model in practice. Some samples in SARD exist solely to trigger vulnerabilities and may not have any practical functionality. Such data evidently cannot help the model improve its performance on real-world data~\cite{chakraborty2021deep}.

\textbf{Limitation 2}: The goal of inserting vulnerabilities into non-vulnerable codebase is to make the generated vulnerabilities more realistic. However, such pattern-based injection limits the types of vulnerabilities that can be generated. Additionally, such injections do not consider if the injected code will affect the semantics of the original codes, which may cause the loss or compromise of the original vulnerability characteristics, 
 thereby generating mislabeled data for model training.



To address these limitations, we propose a natural semantics-preserving transformation method aiming to generate various realistic samples for expanding the vulnerability dataset. This method has two advantages: 1) We formulate transformation rules based on the code style of programmers in the real world, ensuring our generated samples are similar to common real-world coding styles. This guarantees that models trained with our augmented dataset achieve consistent performance in vulnerability detection tasks on real-world datasets. 2) Our transformations will retain the semantics of the original code. Thus, it will not break the context of the vulnerability, which will be able to include vulnerability types present in the original dataset and make the labels of our augmented dataset accurate.

\section{Methodology}
\subsection{Overview}

\begin{figure}[t]
\centering
\vspace{0mm}
\includegraphics[width=1\linewidth]
{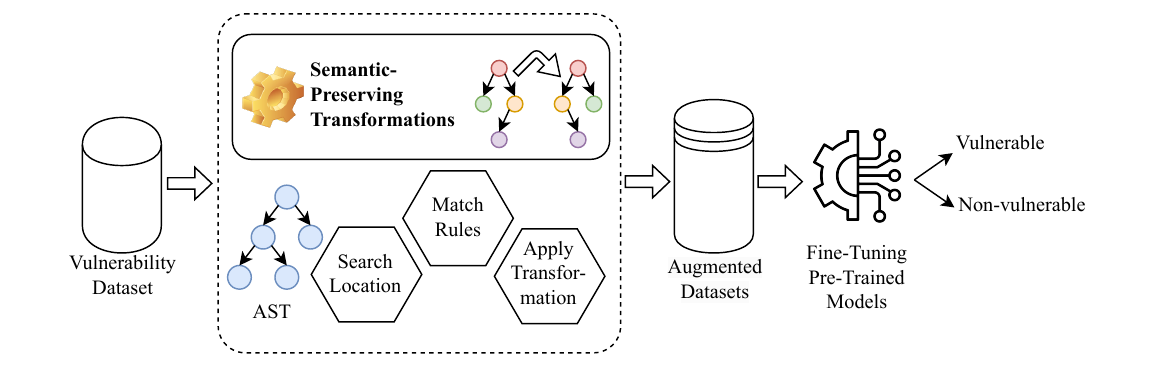}
\vspace{-4mm}
\caption{Overview of semantic-preserving data augmentation}
\vspace{0mm}
\label{fig:overview}
\end{figure}

As shown in ~\autoref{fig:overview}, we provide an overview of our semantic-preserving data augmentation method to facilitate more effective vulnerability detection using pre-trained models. For each sample in the original vulnerability dataset, we first extract its AST and identify potential transformation locations. Next, we examine the applicable rules for all the potential transformation locations in a one-to-one manner, and generate new samples sequentially. We repeat this process for all samples with multiple potential transformation locations to generate as many diverse new samples as possible. Then, the generated data along with the original data together as the augmented dataset is used for fine-tuning the pre-trained language models. 

We will discuss the formulation of the transformations in Section 3.2, introduce the details of the transformation rules in Section 3.3, and present the transformation applying pipeline in Section 3.4.


\subsection{Augmentation Formulation}

Vulnerability detection is a supervised learning binary classification task, where the goal is to predict if a given sample is vulnerable or not. Therefore, the vulnerability dataset typically includes both vulnerability and non-vulnerability samples. As discussed in Section 2.3, the scarcity of real-world vulnerability samples motivates us to perform data augmentation. Besides the challenges of collecting and labeling vulnerability samples, it is also difficult to collect cleaned (noise-free) non-vulnerability samples. After labeling vulnerability samples based on the CVE records, we cannot simply regard the remaining samples without corresponding CVEs as non-vulnerable. This is because not all vulnerabilities are publicly reported to CVE, and there are a certain number of silent vulnerabilities~\cite{wang2019detecting,wang2023graphspd}. Since accurately labeling non-vulnerable data also requires manual inspection, there is also a shortage of cleaned non-vulnerability samples. 

Considering this, our augmentation method should be applicable to not only vulnerable data but also non-vulnerable ones. When designing our method for data augmentation for vulnerability detection, we follow two requirements:

\begin{itemize}
\item \textbf{Requirement 1}: Our method must preserve the original vulnerability and non-vulnerability semantics in the samples. We need to retain the original data's vulnerability and non-vulnerability characteristics while avoiding invalidating the original vulnerabilities or introducing new ones.


\item \textbf{Requirement 2}: Our method needs to make sufficient changes to enrich code representation diversity for better model learning. These changes should also be natural to closely mirror real-world vulnerability and non-vulnerability characteristics.
\end{itemize}
We achieve this through the concept of semantic-preserving transformations using our proposed natural rules. Firstly, we formulate the requirements for the first aspect. Assume we have a mapping $\mathpzc{S}$ that maps code snippets and their inputs to the value changes made in the code and the side effects of these changes. For code snippets $\mathpzc{s}$ and $\mathpzc{t}$, if the input domain of both $\mathpzc{s}$ and $\mathpzc{t}$ is $\mathpzc{I}$, and for any $\mathpzc{i}$ belonging to $\mathpzc{I}$, $\mathpzc{S(s, i)}$ is always equal to $\mathpzc{S(t, i)}$, then we consider $\mathpzc{s}$ and $\mathpzc{t}$ to be semantically equivalent. Assume we have a transformation $\mathpzc{T}$ that applies to a code snippet $\mathpzc{c}$, such that $\mathpzc{S(T(c), \cdot)}$ is always equal to $\mathpzc{S(c, \cdot)}$. In this case, we refer to $\mathpzc{T}$ as a semantics-preserving transformation. Since we strictly control the changes to inputs and their side effects in this process, we believe that this transformation will neither introduce new vulnerabilities nor lose existing ones. Secondly, assuming in the learning-based vulnerability detection task $\mathpzc{D}$, the models use mapping $\mathpzc{M_{C\rightarrow V}^D}(\cdot)$ to map code to vectors to extract features. Since some popular pre-trained code models will use graph-based code representations as supplementary information for their pre-training or fine-tuning, we not only want our method to modify the tokens of the code snippet but also to change the graph structure of the code. Therefore, assuming $\mathpzc{M_{C\rightarrow G}^D}(\cdot)$ is a subprocess of $\mathpzc{M_{C\rightarrow V}^D}(\cdot)$, representing the mapping from code snippets to graph representations, we want our transformation method to also achieve some changes in $\mathpzc{M_{C\rightarrow G}^D}(\cdot)$. In summary, we derive the principles that our transformation design should follow:

\begin{equation} \label{eq:formulation}
    \mathpzc{M^D_{C\rightarrow G}(c) \neq M^D_{C\rightarrow G}(T(c))}
\end{equation}
\begin{equation} \label{eq:verify_formulation}
    \mathpzc{S_F(c,\cdot)}\equiv\mathpzc{S_F(T(c),\cdot)}
\end{equation}

\subsection{Transformation Rules}
\begin{figure}[t]
\centering
\vspace{0mm}
\includegraphics[width=0.9\linewidth, trim=1cm 0cm 1cm 0cm]
{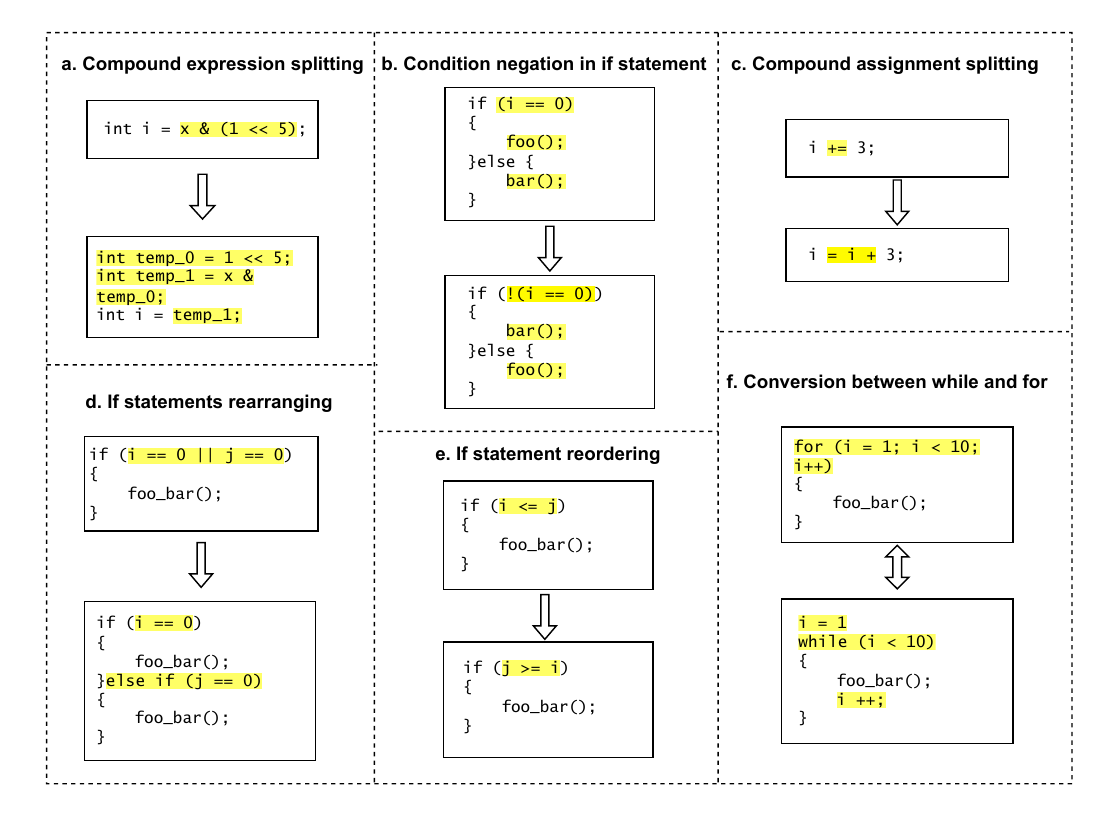}
\vspace{-2mm}
\caption{Example of natural semantic-preserving transformation rules.}
\vspace{0mm}
\label{fig:rules}
\end{figure}

Based on observations of human coding practices in practice, we propose the following types of natural semantic-preserving code transformations to expand existing vulnerability datasets. Examples for each rule are shown in \autoref{fig:rules}.

\begin{enumerate}[a.]
    \item \textbf{Compound expression splitting}: We find that some programmers combine multiple operators and operands to form a single, more complex expression, while others prefer a simpler and more straightforward way of writing expressions. Inspired by this observation, we decompose compound expressions into a series of step-by-step, single-line expressions.
    \item \textbf{Condition Negation in \texttt{if} statement}: In an \texttt{if-else} statement, reversing the logical condition and swapping the two code blocks that are executed is equivalent to the original \texttt{if-else} statement. The choice of which logical form to use depends on the programmer's understanding of the related code logic. Therefore, we can use this rule to transform \texttt{if-else} statements.  
    \item \textbf{Compound assignment splitting}: C/C++ language allows programmers to combine assignment operators with calculating operators, using the compound operators such as \texttt{+=}  and \texttt{/=}  to perform both operations and assignments simultaneously. Some C/C++ programmers who migrated from other languages may not be accustomed to using these shorthand forms of statements and we can observe that both coding styles exist in open-source projects. In this case, we split the specific compound assignment operators into an assignment and an operator on the same line. 
    \item \textbf{\texttt{if} statements rearranging}: Similar to compound expressions, some programmers prefer to use complex logical expressions in \texttt{if} conditions to handle conditions uniformly, while others prefer to use simple logical expressions to handle individual conditions separately. Therefore, we transform complex expressions in \texttt{if} conditions into simple logical expressions to make the transformation. 
    \item \textbf{\texttt{if} statement reordering}: Logical expressions have some equivalent forms (i.e. a \textgreater b is equivalent to b \textless a). We perform transformations by changing these logical expressions in the if condition.
    \item \textbf{Conversion between \texttt{while} and \texttt{for}}: \texttt{while} and \texttt{for} are logically equivalent, so we can convert between these two loop forms by transforming the components of a \texttt{for} statement into statements within multi-line statements in loop body, or by converting the control statements within the body of a while loop into the components of a \texttt{for} statement.

\end{enumerate}

Note that, in a given code sample, there may be multiple locations that match one or more transformation rules. 

\subsection{Applying Transformations}
We perform four steps to apply our natural semantic-preserving transformations on the vulnerability dataset. First, we identify candidate locations in the code where transformations can be applied. Next, we match the rules with the structure of the code locations' context to select the rules to be used. Then, we transform the elements of the code according to these rules. Finally, this process is repeated to apply each rule to multiple matching locations and to apply multiple rules in their respective matching locations.

\begin{center}
\begin{lstlisting}[caption={An example of applying transformations}, label={lst:vulnerable_code}, belowskip=-0.0in, style=mystyle, xleftmargin=.08\textwidth]
void get_size (guint8 * ptr, guint8 ** end)
{
  int cnt = 4;
  int len = 0;
  (*@\hlcyan{while (}@*)(*@\hlcsec{cnt -= 1 }@*)(*@\hlcyan{ >= 0)}@*) {
    int c = *ptr;
    (*@\hlcyan{ptr++;}@*)
    len = (*@\hlcyan{(len {<<} 7) | (c \& 0x7f);}@*)
    (*@\hlcyan{if (!(c \& 0x80))}@*)
      break;
  }
  (*@\hlcyan{if (end)}@*)
    *end = ptr;
  return len;
}
\end{lstlisting}
\end{center}

\vspace{3mm}
\noindent\textbf{Identifying locations.} Since vulnerability datasets are generally collected from real-world projects, the code styles and structures are quite diverse, the location to apply the transformation in each code snippet may vary. Applying rules in some incorrect locations may change the code logic or semantics, and even disrupt the correct syntax structure. Therefore, we generate the corresponding Abstract Syntax Tree (AST) of the code in the first processing step to avoid the impact of different code forms on the identification of transformation locations. Then, we perform pattern searching on the AST to identify transformation locations and check these locations to exclude any potential abnormal sites.

\vspace{3mm}
\noindent\textbf{Matching rules.} After a valid transformation location is identified, the transformations that can be applied to that point will be identified and applied based on our proposed rules in \autoref{fig:rules}. If one rule is able to be applied to multiple locations, we will apply it at each location in a one-to-one manner, generating multiple output samples. Simultaneously, if one location matches multiple rules, we will apply them independently. For example, in the case of \autoref{lst:vulnerable_code}, line 5 will match rule $\mathpzc{c}$ and rule $\mathpzc{f}$. At this point, we will generate two new samples based on these two rules respectively.

\vspace{3mm}
\noindent\textbf{Applying transformation.} Once the pairs of locations and rules are determined, we apply the rule at its corresponding location and keep other context code unchanged. 
The above process describes the application process of transformation for a single rule, and these processing steps are independent of each other. For the transformations involving multiple locations and multiple rules, we will repeat the processing steps for the single transformation multiple times accordingly.
In line 5 of \autoref{lst:vulnerable_code}, after we transform the \texttt{while} loop to a \texttt{for} loop, the statement \texttt{cnt -= 1} will be preserved. At this point, the multi-transformation mechanism will continue to attempt transformations on potentially legal locations in the code, transform \texttt{cnt -= 1} in the new \texttt{for} loop to \texttt{cnt = cnt - 1}, and check to avoid redundancy.

\section{Experimental Evaluation}
\subsection{Experiment Settings}
\vspace{1mm}
\noindent\textbf{Environment.} Our experiments are performed on an Ubuntu 22.04 platform with an Intel W5-2455X, 64 GB RAM, and an NVIDIA A6000 GPU. We use LLVM  18.0.0 to parse the source code and generate the AST for program transformation. 

\vspace{3mm}
\noindent\textbf{Dataset.}
We apply our proposed data augmentation on Big-Vul~\cite{fan2020ac}, a widely adopted function-level vulnerability dataset, to assess if the augmented samples can improve the performance of the given vulnerability detection models. Big-Vul comprises 10,900 vulnerable samples in C/C++ primarily sourced from CVE, spanning 91 Common Weakness Enumerations (CWEs), and originating from over 300 popular open-source projects on GitHub. Additionally, it includes 177,736 non-vulnerable samples.

We randomly divide the dataset into training, validation, and test sets in an 8:1:1 ratio and remove duplicate data and samples that cannot be properly parsed by LLVM. Then, we perform undersampling on non-vulnerability set to achieve class balance. Note that our augmentation is only conducted on the training set. We do not change the validation and test sets to ensure fair evaluation.


\vspace{3mm}
\noindent\textbf{Models.} We evaluate the effectiveness of our data augmentation method on SOTA pre-trained language models since they present superior performance in vulnerability detection tasks.
We select four representative transformer-based pre-trained language models: CodeBERT~\cite{feng2020codebert}, GraphCodeBERT~\cite{guo2020graphcodebert}, UnixCoder~\cite{guo2022unixcoder}, and PDBERT~\cite{liu2024pre}, which have demonstrated excellent performance when fine-tuned for downstream vulnerability detection tasks ~\cite{zhang2022reentrancy, wang2024empirical, ni2023function,nie2023understanding }.  

\begin{itemize}
    \item \textbf{CodeBERT} is the first bi-modal transformer model pre-trained with natural language and program language pairs from various repositories on GitHub. 

\vspace{1mm}
\item \textbf{GraphCodeBERT} incorporates data flow graph (DFG) structural information to capture relationships between code elements instead of the sequential approach used by CodeBERT.

\vspace{1mm}
\item \textbf{UnixCoder} adopts an encoder-decoder architecture and leverages AST information with code comments during its pre-training phase to enhance code representation.

\vspace{1mm}
\item \textbf{PDBERT} utilizes two pre-training objectives, Control Dependency Prediction (CDP) and Data Dependency Prediction (DDP), to guide the model for learning the knowledge required for analyzing fine-grained code semantics.
\end{itemize}

\noindent We follow previous works~\cite{shi2022compressing, ni2023function} to fine-tune GraphCodeBERT and UnixCoder. CodeXGLUE framework~\cite{lu2021codexglue} is used to fine-tune and test CodeBERT~\cite{feng2020codebert}. We adopt the same settings and implementation as PDBERT's original work~\cite{liu2024pre}.


\vspace{3mm}
\noindent\textbf{Evaluation Metrics}: We use the Accuracy, Precision, Recall, and F1 score to evaluate the improvement of fine-tuning pre-trained models for vulnerability detection with the augmented dataset (Section 4.3). We also use them in the comparison evaluation with existing data augmentation works (Section 4.4). When further checking the performance improvement by vulnerability type (Section 4.5), we examine the true positive rate (TPR) for each specific type.

\subsection{Implementation}
\label{sec:Implementation}

We apply the transformations introduced in Section 3.2 on the entire Big-Vul dataset and generate 3,233,635 new samples in total. We randomly select 100 generated vulnerability samples and 100 non-vulnerability samples and manually check the quality and correctness of their code. We found that all of them can be parsed by Clang, and none introduce syntax errors or semantic changes. 

Among them, 2,883,706 (89.2\%) samples are generated after performing multiple transformation rules on multiple locations. This means that their original samples were generated by applying multiple transformation rules at various locations, indicating that their original samples had more than one location suitable for several transformation rules.
The widest applicable rule is the negation of the condition in \texttt{if} statements, covering 63.8\% of all cases. The following one is the \texttt{if} statement reordering, which accounts for 19.0\%. 

\subsection{Vulnerability Detection Improvement to Pre-Trained Models}
\label{sec:Improvement}
Here, we evaluate if using data augmentation on the vulnerability dataset for fine-tuning the four representative pre-trained language models can improve their performance on vulnerability detection.


\begin{table*}[tp]
\renewcommand{\arraystretch}{1.05}
\begin{center}
\setlength{\tabcolsep}{2mm}
\caption{Performance improvement using augmented datasets}
\label{tab:Improvement}
\footnotesize	
\begin{tabular}{c|c|c c c c}
\toprule

    \textbf{Model} & \textbf{Dataset} & \textbf{Accuracy} & \textbf{Precision} & \textbf{Recall} & \textbf{F1} \\
    \hline\hline
    \multirow{3}{*}{CodeBERT}          & original  & 66.8$\%$ & 76.8$\%$ & 48.0$\%$ & 59.1$\%$ \\
    \multirow{3}{*}{}                  & augm-vul   & 70.7$\%$ & \textbf{84.0$\%$} & 51.2$\%$ & 63.6$\%$ \\
    \multirow{3}{*}{}                  & augm-both & \textbf{75.5$\%$} & 77.5$\%$ & \textbf{71.9$\%$} & \textbf{74.6$\%$} \\
\hline
    \multirow{3}{*}{GraphCodeBERT}     & original  & 63.2$\%$ & 82.1$\%$ & 33.8$\%$ & 47.8$\%$ \\
    \multirow{3}{*}{}                  & augm-vul   & 66.0$\%$ & \textbf{88.4$\%$} & 36.8$\%$ & 52.0$\%$ \\
    \multirow{3}{*}{}                  & augm-both & \textbf{73.3$\%$} & 76.9$\%$ & \textbf{66.7$\%$} & \textbf{71.4$\%$}  \\
\hline
    \multirow{3}{*}{UnixCoder}         & original  & 74.7$\%$ & 79.5$\%$ & 66.6$\%$ & 72.4$\%$ \\
    \multirow{3}{*}{}                  & augm-vul   & 71.6$\%$ & \textbf{87.6$\%$} & 50.4$\%$ & 64.0$\%$ \\
    \multirow{3}{*}{}                  & augm-both & \textbf{75.1$\%$} & 79.8$\%$ & \textbf{67.1$\%$} & \textbf{72.9$\%$} \\
\hline
    \multirow{3}{*}{PDBERT}            & original  & 73.7$\%$ & 79.7$\%$ & 63.5$\%$ & 70.7$\%$ \\
    \multirow{3}{*}{}                  & augm-vul   & \textbf{74.7$\%$} & \textbf{81.7$\%$} & \textbf{63.7$\%$} & \textbf{71.6$\%$} \\
    \multirow{3}{*}{}                  & augm-both & 73.8$\%$ & 81.7$\%$ & 61.3$\%$ & 70.1$\%$ \\
\hline
\bottomrule
\end{tabular}
\end{center}
\end{table*}

\vspace{3mm}
\noindent\textbf{Augmentation Setting.} Since collecting accurately labeled vulnerability and non-vulnerability samples are both challenging, we evaluate two settings for augmenting the given vulnerability dataset:

\begin{itemize}
    \item \textbf{Augmenting vulnerable data only (augm-vul).} Considering the scarcity of vulnerabilities in the real world, we only augment the vulnerable samples in the training set and obtain 22,893 vulnerable samples from 7,631 original Big-Vul vulnerabilities in the training set. We randomly select an equal number of non-vulnerability samples from the original Big-Vul for balancing the classes. In this way, we increase the size of the original training set by 300\%.
    
    \vspace{1mm}\item \textbf{Augmenting both vulnerable and non-vulnerable data (augm-both).} Since existing noise-free non-vulnerability data is also limited, they may not well represent the real-world data characteristics either. Therefore, we perform augmentation on both vulnerable and non-vulnerable data and get 22,893 vulnerable samples and 22,893 non-vulnerable samples after undersampling, which also expands the original training set by 300\%.
\end{itemize}


\noindent\textbf{Performance Improvement.} The test results with our augmented data are shown in \autoref{tab:Improvement}. The ``original" refers to the results using the original Big-Vul training set. ``augm-vul" and ``augm-both" are corresponding to the results under the above-mentioned two augmentation settings. 
We can see that across all four pre-trained models, adopting our augmented dataset presents noteworthy improvements compared with the original set. 
In CodeBERT and GraphCodeBERT, the accuracy of the models respectively increases 8.7$\%$ and 10.1$\%$, and the F1 increases  15.5$\%$ and 23.6$\%$. 
In UnixCoder and PDBERT, our augmented datasets also exhibit improvements, albeit less than the first two models. This may be attributed to structural differences among the models: CodeBERT and GraphCodeBERT do not sufficiently learn control dependency information during pre-training. As our transformation rules primarily introduce variations in control dependencies, the resulting improvements on these two models are greater compared to UnixCoder and PDBERT, which already incorporate control dependencies during their pre-training.

For most models, augmenting both vulnerable and non-vulnerable datasets achieves higher performance. The reason is that the original dataset lacks a complete representation of real-world data for both vulnerable and non-vulnerable cases. With the help of our proposed data augmentation method, the generated data enriches both vulnerability and non-vulnerability data diversity and variety, thereby improving generalization results. In PDBERT, augmenting vulnerabilities only performs slightly better. This may be because PDBERT incorporates additional program dependency information during pre-training, which already captures certain characteristics provided by our augmented dataset.

In summary, our data augmentation approach using natural semantic-preserving transformation can greatly help improve the performance of pre-trained language models. In most cases, augmenting both vulnerable and non-vulnerable data in the vulnerability dataset yields the best detection results. 


\subsection{Comparison with Other Vulnerability Dataset Augmentation Methods}

\begin{table*}[tp]
\renewcommand{\arraystretch}{1.05}
\vspace{-0mm}
\begin{center}
\setlength{\tabcolsep}{2mm}
\caption{Comparison with other vulnerability dataset augmentation methods}
\vspace{-0mm}
\label{tab:Comparison}
\footnotesize	
\begin{tabular}{c|c c}
\toprule

  \textbf{Dataset} & \textbf{Accuracy} & \textbf{F1} \\
    \hline\hline
    Original  & 59.1$\%$   & 46.6$\%$ \\
    \hline
    SARD (augm-both)         & 58.7$\%$ (0.7$\%\downarrow$)   & 53.3$\%$ (14.4$\%\uparrow$) \\
    \hline
    VGX (augm-vul)         & 60.6$\%$ (2.5$\%\uparrow$)   & 55.8$\%$ (19.7$\%\uparrow$) \\
    \hline
    Ours (augm-both)     & \textbf{67.2$\%$ (13.9$\%\uparrow$)}  & \textbf{65.2$\%$ (39.9$\%\uparrow$)}  \\

\bottomrule
\end{tabular}
\end{center}
\end{table*}

To demonstrate the advantage of our method, we conduct comparison experiments using augmented data to fine-tune GraphCodeBERT for vulnerability detection tasks. In our evaluation, we choose two representative vulnerability augmentation approaches: 

\begin{itemize}
    \item \textbf{SARD}~\cite{SARD} is a well-known publicly available dataset, mainly composed of synthetic data. It performs program transformation on both vulnerability and non-vulnerability samples by adding always-true conditions or dummy statements that will not affect the original functionalities. Although its program transformation tool is not open source, we can directly select vulnerability and non-vulnerability samples from SARD to add to our training set. This is because SARD is built using real-world popular OSS, the similar source to our adopted Big-Vul dataset. 
    

    \vspace{1mm}\item \textbf{VGX}~\cite{nong2024vgx} is a SOTA injection-based vulnerability augmentation method, which extracts real-world vulnerabilities and applies it to non-vulnerable codebase. We directly use VGX's publicly released augmented dataset, which includes vulnerability samples that are generated by extracting functions from projects included in Big-Vul for vulnerability injection, as well as real-world non-vulnerable samples. 
    Since VGX cannot augment non-vulnerable samples, we regard it as an augm-vul setting.
\end{itemize}

It is important to note that SARD also includes samples from CVE and will partially overlap with Big-Vul, which may lead to data leakage. To address this issue, we split Big-Vul's samples based on their publication dates, ensuring that data from the same CVE as SARD does not appear in the test set.
Specifically, we use 2017 as the cutoff and divide Big-Vul into a training set of 16,110 samples and validation/test sets of 1,990 samples. Then, following the setting in \ref{sec:Improvement}, we augment the training set by separately adding SARD, VGX, and our transformed samples, expanding each to 300\% of its original size.

As shown in~\autoref{tab:Comparison},
compared with SARD and VGX, our method achieves the best improvement, increasing the accuracy by 13.9\% and the F1 score by 39.9\% over the original training set. The performance improvement by VGX is minimal. This may be because some vulnerability injections alter the original semantics of the code, thus failing to ensure the vulnerability characteristics of the generated code.
We can observe that fine-tuning with SARD even makes the model performance worse. Although SARD employs semantic-preserving transformations, it does not guarantee the naturalness of the code, resulting in differences between its generated code and real-world code. This makes training models with SARD less effective in improving test performance on real-world datasets. 

Therefore, we conclude that our method can generate natural vulnerability and non-vulnerability samples that more closely resemble real-world data distribution. This leads to higher improvements in vulnerability detection using pre-trained models compared to other data augmentation approaches.

\subsection{Vulnerability Detection Improvement by Type}

In this subsection, we further investigate the detection improvements facilitated by our data augmentation for each vulnerability type. 

\autoref{tab:CWEs} presents the performance improvement of GraphCodeBERT on top 25 frequent CWEs collected from CVE by Big-Vul, which reflects the vulnerability distribution in the real world. The first column lists the CWE IDs. The second column shows the number and proportion of each CWE in the training set. The third column displays the number of test samples and the column named ``TPR Impr" refers to the TPR improvement from the original dataset and augmented dataset under augm-both setting for each CWE.


After fine-tuning with our augmented dataset, we observe TPR improvement in 23 out of the 25 CWEs, with an average increase of over 30\%. The only exceptions, CWE-772 and CWE-269, have very few test samples (6 and 3, respectively), which may cause fluctuations in the test results. Thus, we determine that fine-tuning pre-trained models with our augmented data significantly enhances their ability to detect prevalent vulnerabilities in practice.



\subsection{Overhead Evaluation}
The overhead of data augmentation in our work is composed of two parts: 1) the time used by LLVM to parse the input code and generate corresponding ASTs, which cannot be controlled by us; and 2) the time to analyze the generated ASTs and perform program transformation to generate augmented data. This process is handled by Python scripts totaling 7,000 lines of code developed by us.

Including the above two parts, it takes 33 hours in total to generate 3,233,635 new samples by transforming all 188,636 vulnerable and non-vulnerable samples of the entire Big-Vul dataset. In other words, an average of approximately 28 samples can be generated per second, which is acceptable. 


\begin{table*}[tp]
\renewcommand{\arraystretch}{1.25}
\setlength{\tabcolsep}{0pt}
{
\scriptsize
\vspace{-0mm}
\begin{center}
\setlength{\tabcolsep}{2mm}
\caption{Performance improvement of GraphCodeBERT on top 25 CWEs}
\vspace{-0mm}
\label{tab:CWEs}	
\begin{tabular}{p{0.65cm}<{\centering}|p{1.55cm}<{\centering}| p{0.4cm}<{\centering} |p{1.55cm}<{\centering} | l}

\toprule

    \textbf{CWE} & \textbf{Train(\%)} & \textbf{Test} & \textbf{TPR Impr}  &  \textbf{CWE Description} \\
    \hline\hline
    119 & 1,526(20.0\%) & 179 & 47.5\%$\rightarrow$70.4\% & Improper Op Restriction in Mem Buffer Bounds \\
    20 & 812(10.6\%)  & 100 & 40.0\%$\rightarrow$73.0\% & Improper Input Validation \\
    399 & 529(6.9\%)  & 86 & 23.3\%$\rightarrow$58.1\% & Resource Management Errors \\
    264 & 365(4.8\%)  & 49 & 22.4\%$\rightarrow$61.2\% & Permissions, Privileges, and Access Controls
 \\
    416 & 225(2.9\%)  & 33 & 18.2\%$\rightarrow$60.6\% & Use After Free \\
    200 & 321(4.2\%)  & 39 & 38.5\%$\rightarrow$61.5\% &  Sensitive Info Exposure to Unauthorized Actor \\
    125 & 367(4.8\%)  & 55 & 47.3\%$\rightarrow$65.5\% & Out-of-bounds Read \\
    189 & 253(3.3\%)  & 26 & 34.6\%$\rightarrow$73.1\% & Numeric Errors \\
    362 & 202(2.6\%)  & 23 & 39.1\%$\rightarrow$69.6\% & Race Condition \\
    476 & 136(1.8\%)  & 18 & 33.3\%$\rightarrow$66.7\% & NULL Pointer Dereference \\
    190 & 146(1.9\%)  & 13 & 30.8\%$\rightarrow$76.9\% & Integer Overflow or Wraparound \\
    254 & 87(1.1\%)  & 18 & 11.1\%$\rightarrow$61.1\% & Security Features \\
    787 & 124(1.6\%)  & 14 & 42.9\%$\rightarrow$71.4\% & Out-of-bounds Write \\
    284 & 136(1.8\%)  & 16 & 12.5\%$\rightarrow$43.8\% &  Improper Access Control \\
    732 & 44(0.5\%)  & 4 & 0\%$\rightarrow$50.0\% & Incorrect Permission Assign for Critical Resource \\
    310 & 70(0.9\%)  & 10 & 60.0\%$\rightarrow$100\% &  Cryptographic Issues \\
    400 & 28(0.4\%)  & 4 & 25.0\%$\rightarrow$75.0\% & Uncontrolled Resource Consumption \\
    772 & 36(0.5\%)  & 6 & 100\%$\rightarrow$66.7\% & Missing Resource Release after Effective Lifetime \\
    59 & 34(0.4\%)  & 9 & 11.1\%$\rightarrow$66.7\% & Improper Link Resolution Before File Access \\
    617 & 18(0.2\%)  &  0 & - & Reachable Assertion  \\
    415 & 56(0.7\%)  &  7 & 57.1\%$\rightarrow$85.7\% & Double Free \\
    269 & 22(0.3\%)  & 3 & 33.3\%$\rightarrow$33.3\% & Improper Privilege Management \\
    404 & 45(0.6\%)  & 3 & 0\%$\rightarrow$66.7\% & Improper Resource Shutdown or Release \\
    704 & 20(0.3\%)  & 1 & 0\%$\rightarrow$100\% & Incorrect Type Conversion or Cast \\
    79 & 43(0.3\%) & 5 & 20.0\%$\rightarrow$40.0\% & Cross-site Scripting \\
    Others & 1986(26.1\%)  & 233 & 26.2\%$\rightarrow$66.5\% & Other CWEs \\ \hline
   All & 7,631(100\%) & 954 & 33.8\%$\rightarrow$66.7\% & All CWEs \\

\hline
\bottomrule
\end{tabular}
\end{center}}
\end{table*}

\section{Discussion}


Like most vulnerability detection works, we focus on C/C++ since they are the programming languages with the highest number of vulnerabilities.
It is worth noting that while examples of our transformation rules are specifically presented in C/C++, our method, based on LLVM parsing, can be easily extended to other languages supported by LLVM, such as Rust. 

In Section~\ref{sec:Improvement}, we observe that the improvements are not very significant when fine-tuning UniXcoder and PDBERT with our augmented dataset. We speculate that this is because these two models have already used a large amount of graph-based code representation data during the pre-training phase, which enables them to effectively capture various code features. To further enhance their performance, we may need to generate data that exhibits greater diversity. We leave the generation of more complex data for our future work.

Currently, our seven transformations are manually summarized, and their application framework is specified for each transformation based on our observation. 
In recent years, deep learning-based models have shown great potential in code translation. We will explore the potential of automatically performing semantics-preserving transformations with these techniques in our future work.

Our work performs semantic-equivalent program transformation on existing known vulnerable code. All experiments are conducted on publicly available vulnerability datasets and vulnerability detection models. Since no new vulnerabilities will be generated or identified by this work, no ethical issues will arise.


\section{Related Work}
\subsection{Code Editing}
The code editing task aims to automatically edit a code snippet and produce a modified new version of the code. Some methods~\cite{tufano2019learning, tufano2021towards} train seq2seq models directly using pull request data from open-source projects to perform code changes, while other methods~\cite{chakraborty2020codit} extract and modify the code's AST to ensure syntactic correctness. Subsequent works~\cite{wang2021codet5, guo2022unixcoder} introduce the transformer architecture. Some approaches, like NatGen~\cite{chakraborty2022natgen}, focus more on the naturalness of the edited code. However, because these methods cannot guarantee semantic consistency before and after code modifications, they cannot ensure that the edited code still contains vulnerabilities. This makes these approaches unsuitable for augmenting vulnerability datasets.

\subsection{Vulnerability Generation}
The mainstream approach to augmenting vulnerability datasets involves injecting or synthesizing new vulnerabilities based on existing code. FixReverter~\cite{zhang2022fixreverter} uses artifact patterns to inject vulnerability code for the fuzzing tools evaluation. Fuzzle~\cite{lee2022fuzzle} treats function call chains as a maze and synthesizes new vulnerable programs by generating new mazes. VulGen~\cite{nong2023vulgen} trains a transformer-based model to address the location to inject the vulnerability. VGX ~\cite{nong2024vgx} leverages code semantics-aware Transformer attention and human knowledge to find out where and how to inject the vulnerabilities. Liu et al.~\cite{liu2024enhancing} use vulnerability-preserved transformations to augment the dataset for the training of a GGNN model. 
However, the above approaches either only support generating specific types of vulnerabilities, and increasing the variety of vulnerability types requires a substantial manual workload; or they cannot ensure that the generated code keeps correct syntax, which may mislead downstream task models during training and subsequently impact the final task performance. 
By using semantics-preserving transformation methods, we can easily generate a large number of syntactically correct and natural vulnerability samples. 
\section{Conclusion}

It is imperative to timely and effectively detect software vulnerabilities in modern communication network systems. This paper proposes a method to augment vulnerability detection datasets by using natural semantics-preserving transformations to enhance the performance of SOTA pre-trained language models for vulnerability detection. Through experimental evaluation, we find that our method can generate a large number of samples with enriched data diversity and variety. We evaluate our generated data on four representative code pre-trained models and get up to 10.1$\%$ improvement in accuracy and 23.6$\%$
increase in F1. Also, our augmentation method greatly outperforms other representative augmentation methods in enhancing pre-trained models in vulnerability detection.

\section{Acknowledgement}

We are grateful for the opportunities provided by the UTD Summer Research Program for high school students. Jiahao Cao's research is supported by the NSFC Grant 62202260.

%
%
%
%

\bibliographystyle{acm}
\bibliography{paper}

\end{document}